\documentclass[useAMS,usenatbib]{mn2e} %,usenatbib
\usepackage[english]{babel}
\usepackage[T1]{fontenc}
\usepackage[latin9]{inputenc}
\usepackage{xcolor}
\usepackage{float}
\usepackage{textcomp}
\usepackage{graphicx}
\usepackage{amssymb}

\providecommand{\tabularnewline}{\\}

\def\apjl{ApJL }

\def\apj{ApJ }

\def\aap{A\&A }
\def\nat{Nature }
\def\mnras{MNRAS }
\def\ssr{Space~Sci.~Rev. }
\def\apss{Ap\&SS }
\def\pasj{PASJ }
\def\actaa{Acta Astron. }
\usepackage{hyperref}
\hypersetup{colorlinks=true,citecolor=blue,linkcolor=black,filecolor=black,runcolor=black}

\title[SGR 1806$-$20 distance and dust properties in molecular clouds]{SGR 1806$-$20 distance and dust properties in molecular clouds by
analysis of a flare x-ray echoes}

\author[G. Svirski et al.]{Gilad Svirski$^{1}$, Ehud Nakar$^{1}$ and Eran O. Ofek$^{2,3}$\\
$^{1}$Raymond and Beverly Sackler School of Physics \& Astronomy, Tel Aviv University, Tel Aviv 69978, Israel\\
$^{2}$Division of Physics, Mathematics and Astronomy, California Institute
of Technology, Pasadena, CA 91125, USA\\
$^{3}$Einstein Fellow}

\begin{document}
\maketitle
\begin{abstract}
The soft gamma repeater SGR 1806\textminus{}20 is most famous for
its giant flare from 2004, which yielded the highest $\gamma$-ray
flux ever observed on Earth. The flare emphasized the
importance of determining the distance to the SGR, thus revealing
the flare's energy output, with implications on SGRs energy budget
and giant flare rates. We analyze x-ray scattering echoes observed
by ${\it Swift}$/XRT following the 2006 August 6 intermediate burst of
SGR 1806\textminus{}20. Assuming positions and opacities of the
molecular clouds along the line-of-sight from previous works, we
derive direct constrains on the distance to SGR 1806\textminus{}20,
setting a lower limit of 9.4 kpc and an upper limit of 18.6 kpc
(90\% confidence), compared with a 6\textminus{}15 kpc distance
range by previous works. This distance range matches an energy
output of $\approx10^{46}$ erg$\,$s$^{-1}$ for the 2004 giant
flare. We further use, for the first time, the x-ray echoes in order
to study the dust properties in molecular clouds. Analyzing the
temporal evolution of the observed flux using a dust scattering
model, which assumes a power-law size distribution of the dust
grains, we find a power-law index of $-3.3_{-0.7}^{+0.6}$
($1\sigma$) and a lower limit of $0.1\mu\mbox{m}$ ($2\sigma$) on the
dust maximal grain size, both conforming to measured dust properties
in the diffused interstellar medium (ISM). We advocate future burst
follow-up observations with ${\it Swift}$, $\emph{Chandra}$ and the
planned $\emph{NuSTAR}$ telescopes, as means of obtaining much
superior results from such an analysis.
\end{abstract}
\begin{keywords}
gamma-rays, X-rays: individual: SGR 1806-20, ISM: clouds,
dust, extinction
\end{keywords}

\section{Introduction}\label{sec:Introduction}

Soft Gamma Repeaters (SGRs) are objects emitting
soft gamma-ray and hard x-ray bursts at irregular intervals, as well as a persistent x-ray emission (for a review
see \citealt{2006csxs.book..547W} and references therein). Bursts
are typically short $\left(\sim0.1\,{\rm s}\right)$ and are gathered
within active periods that last between a few weeks to several
months, followed by years of quiescence \citep{1998AAS...193.5602K}.
Bursts are commonly classified according to their peak luminosity,
from the most common flares reaching $10^{41-42}\,$
erg$\,$s$^{-1}$, up to the rare 'giant' flares reaching
$\sim10^{46}\,$ erg$\,$s$^{-1}$. SGRs are believed to be magnetars
\textminus{} neutron stars with a surface magnetic field of
$\sim 10^{15}\,$G, which serves as the energy source of the bursts and the persistent emission
\citep{1992AcA....42..145P,1992ApJ...392L...9D,1998Natur.393..235K}.

SGR 1806$-$20, lying in the direction of the Galactic center behind
a veil of $15-30$ magnitudes of optical extinction \citep[hereafter CE04]{2004A&A...419..191C},
is one of a handful known SGRs. Its resume includes the first SGR
to be observed, on January 7 1979 (originally classified as a gamma ray burst,
\citealt{1981Ap&SS..75...47M}) and the subject of the first SGR spindown
rate measurement \citep{1998Natur.393..235K} $-$ a major milestone
in the acceptance of the magnetar hypothesis. Yet it is most famous
for producing the most energetic giant flare observed to date; on
2004 December 27 it emitted a 0.1$\,$s flare with an estimated total
energy of $2-5\times10^{46}\left(\frac{d}{15\,\mbox{kpc}}\right)^{2}$
erg, where $d$ is the distance to the SGR \citep{2005Natur.434.1107P,2005Natur.434.1098H}.
The emitted energy in this event is evaluated 100 times higher than
the next most energetic SGR recorded event (assuming $d=15$ kpc).

An energy of $E\sim10^{46}\,$erg is at odds with some of the observations.
First, a naive rate of $\frac{1}{150}$ yr$^{-1}$ flares with similar
energy per SGR%
\footnote{based on a single $E\approx10^{46}$ ergs event within 30 years of
observing $\sim5$ SGRs, as of 2004%
} is ruled out by failure to observe corresponding population of
extragalactic SGRs (detectable to $\sim30$ Mpc,
\citealt{2005Natur.434.1107P,2006ApJ...640..849N}). Moreover, even a
rate of a single giant flare per SGR life span, within $2\sigma$ of
the observed rate obtained by a careful statistical treatment, has
only marginal agreement with observed extragalactic rates
\citep{2007ApJ...659..339O}. Second, the surface magnetic field
corresponding to observed properties of SGRs (e.g. spindown rate) is
$B\approx10^{15}G$, matching an external magnetic energy of
$\sim10^{47}\mbox{erg}$, comparable to the energy output of a single
giant flare. Since the source of both the energy reservoir powering
giant flares and the reservoir responsible for the persistent
emission is thought to be the surface magnetic field of the SGR, one
might expect a significant transformation in the observed spectral
and temporal emission parameters of SGR 1806\textminus{}20 following
the giant flare. This expectation is not met by
observations\footnote{These works do report changes in
temporal characteristics of the SGR following the giant flare,
suggesting reorganization of the magnetic field. Similar reports
followed the giant flare of SGR 1900+14, e.g.
\cite{2004NuPhS.132..604G}. However, the loss of a significant
portion of the total energy reservoir seems to justify a more
dramatic change.} \citep{2007ApJ...654..470W,2007A&A...476..321E}.
A possible explanation for both the rates discrepancies and the
unchanged emission features is a distance shorter than the commonly
assumed 15 kpc to SGR 1806\textminus{}20, matching a less energetic
flare output.

Employing different approaches, several papers from recent years suggest
distance ranges within 6\textminus{}15 kpc to SGR 1806\textminus{}20.
The emerging factor of nearly 3 in the SGR's distance estimates translates
to nearly an order of magnitude difference in its emitted energy.
CE04 used CO emission lines and $\mbox{NH}_{3}$ absorption features
from molecular clouds along the line of sight to find the clouds radial
velocities, inferring two possible locations per cloud, one in front
of the Galactic center and one behind it. Accounting for the optical
extinction of the star powering the nebula LBV 1806\textminus{}20
they determined a distance of $15.1_{-1.3}^{+1.8}$ kpc ($2\sigma$)
to the cluster containing LBV 1806\textminus{}20. They associated
SGR 1806\textminus{}20 with this cluster due to its angular proximity
of 12$''$ to LBV 1806\textminus{}20 and the match between SGR 1806\textminus{}20
x-ray absorption and the IR extinction towards the cluster members.
\citet{2004ApJ...610L.109F} measured radial velocities using absorption
lines from LBV 1806\textminus{}20 and nearby stars, which translated
to a distance of 11.8 kpc. \citet{2008MNRAS.386L..23B} spectroscopically
classified several stars which were identified as members of the cluster
of LBV 1806\textminus{}20 by CE04 and \citet{2005ApJ...622L..49F}.
Based on their absolute magnitude calibration and near-IR photometry,
as well as isochrones fit to the cluster's age, they obtained a cluster
distance of $8.7_{-1.5}^{+1.8}$ kpc. As opposed to the above associative
distance estimations, \citet{2005Natur.434.1112C} gave a more direct
estimate. They identified the decaying bright radio afterglow of the
December 2004 giant flare a week after the burst. Based on absorption
features of intervening interstellar neutral hydrogen clouds, they
constrained SGR 1806\textminus{}20 distance to within 6.4\textminus{}9.8
kpc. \citet{2005ApJ...630L.161M} accepted the lower limit of $\sim6$
kpc but rejected their upper limit, disqualifying the association
of the absorption feature used to set this limit with SGR 1806\textminus{}20.

We present a new direct estimate of the distance to SGR 1806\textendash{}20
based on dust scattered x-ray observations, from which we also extract
properties of the dust along the line of sight. The scattering of
x-rays from dust grains in the ISM was first considered by \citet{1965ApJ...141..864O}.
For an x-ray source with a varying intensity (e.g. a short burst),
if the dust spatial distribution is known, one can use the time delay
between the direct signal and the scattered signal to constrain the
distance to the x-ray source \citep{1973A&A....25..445T}. This was
first applied for constraining the distance to the X-ray binary Cyg
X-3 \citep{2000A&A...357L..25P}.

Analysis of x-ray halos around Galactic
and extragalactic sources have been used to constrain the properties
of dust grains in the ISM \citep{1986ApJ...302..371M,1995A&A...293..889P},
with results conforming to the dust model by \citet{1977ApJ...217..425M}.
Such analyses exploit the dependence of the scattering cross section
on the grain properties as well as the dependence of the halo radial
profile on the positions of the scattering grains along the line of
sight.

In the case of both a short burst and a thin dust scatterer, the halo
is replaced by a ring which radius increases with time. Measuring
the expansion rate of such observed rings from GRBs has been used
to derive distances to Galactic dust clouds \citep{2007A&A...473..423V}
as well as the time of the original burst \citep{2010MNRAS.404.1018F}.
\citet{2010ApJ...710..227T} combined observations by both ${\it Swift}$/XRT
and XMM-Newton/EPIC of rings following bursts of the anomalous X-ray
pulsar 1E1547.0\textendash{}5408, and used several different models
for the dust composition and grain size distribution to fit the intensity
decay of each ring as a function of time and energy, in order to obtain
constrains on the distance to the X-ray source. They concluded that
in the absence of independent constraints on the distance of the source
or the scatterer, their analysis is highly sensitive to the size of
the largest dust grains, as reflected by each of the models used.
A dust-scattered halo surrounding an SGR was first reported by \citet{2001ApJ...558L..47K}
for SGR 1900+14, with data quality insufficient for further analysis.

The dust along the line of sight to SGR 1806\textminus{}20 is
concentrated within molecular clouds (e.g. CE04). As opposed to the
diffuse ISM, where dust properties are efficiently probed using UV
and longer wavelengths, the dust in cores of molecular clouds is not
easily accessible due to the clouds high optical depth at these
wavelength. Moreover, observational evidence
\citep{1973ApJ...182...95C,1980ApJ...235...63J,1997ApJ...491..615G,2003A&A...398..551S,2007ApJ...666L..73C,2007ApJ...669..493W,2008ApJ...684.1228S,2009ApJ...696..484B}
and theoretical considerations (e.g.
\citealt{1993A&A...280..617O,1994ApJ...430..713W,2009A&A...502..845O})
indicate that the high density in molecular clouds leads to grain
coagulation that alters the dust grain size distribution from the one
observed in the diffuse ISM. Since molecular clouds are optically
thin to hard x-rays, small angle x-ray scattering provides a unique
tool to directly probe grain size distribution at their cores.
Despite this virtue, x-rays have not been used yet to probe dust
properties in molecular clouds. Here we seize the opportunity to
employ, for the first time, x-ray echoes for this purpose.

On 2006 August 6, an intermediate burst of SGR 1806\textminus{}20
was observed \citep{2006GCN..5416....1H,2006GCN..5419....1H}.
Anticipating delayed dust scattered echoes, ${\it Swift}$/XRT took a
dozen observations during the two weeks following the burst. A
preliminary analysis of the first two observations by
\citet{2006GCN..5438....1G} reported an expanding halo due to dust
scattering. We reanalyze ${\it Swift}$ observations, finding the
halo flux profile of each observation. Assuming properties of
molecular clouds along the line of sight to SGR 1806\textminus{}20
as reported in CE04, we use the observation profiles to constrain
the distance to SGR 1806\textendash{}20. We then assume, instead of
the CE04 dust distribution, a single predominant dust screen along
the line of sight and a power-law distribution for the dust grain
size, and use the observations to constrain the grain size
distribution within the intervening molecular cloud that
dominates the scattering.

This paper is organized as follows. Observations are described in
Section 2. Section 3 reviews the dust scattering model we use in
our analyses, which are presented in Section 4. In section 4.1 we
assume the distance of the scattering clouds is known from CE04 and
constrain the distance to SGR 1806\textminus{}20. In section 4.2 we
relax our assumptions regarding the distance to the scattering clouds
and study the properties of the scattering dust. We discuss implications
of our analyses for future observations in section 5 and draw conclusions
in section 6.

\section{Observations}\label{sec:Observations}

\begin{figure*}
\centering
\includegraphics[scale=0.4]{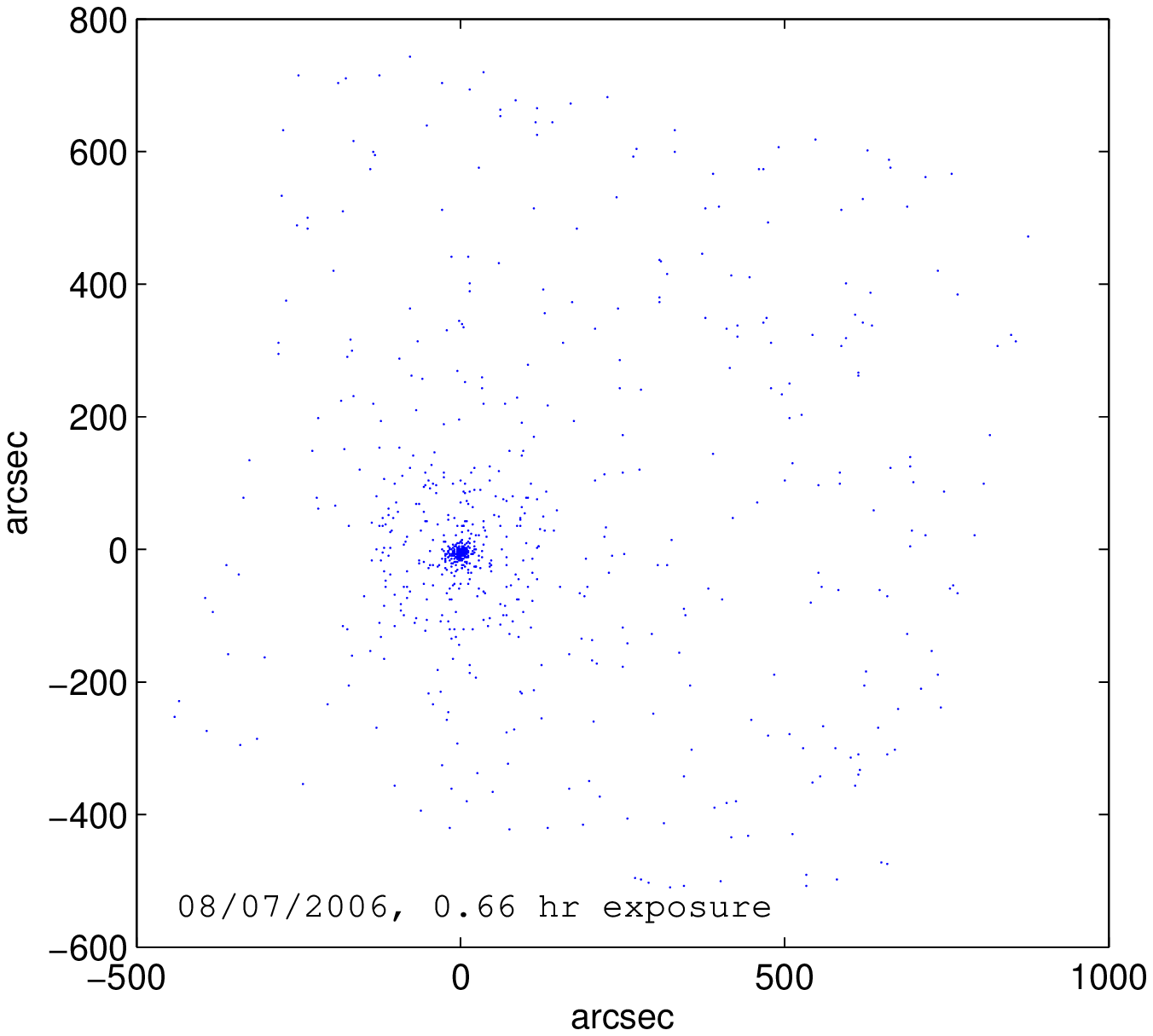}
\includegraphics[scale=0.4]{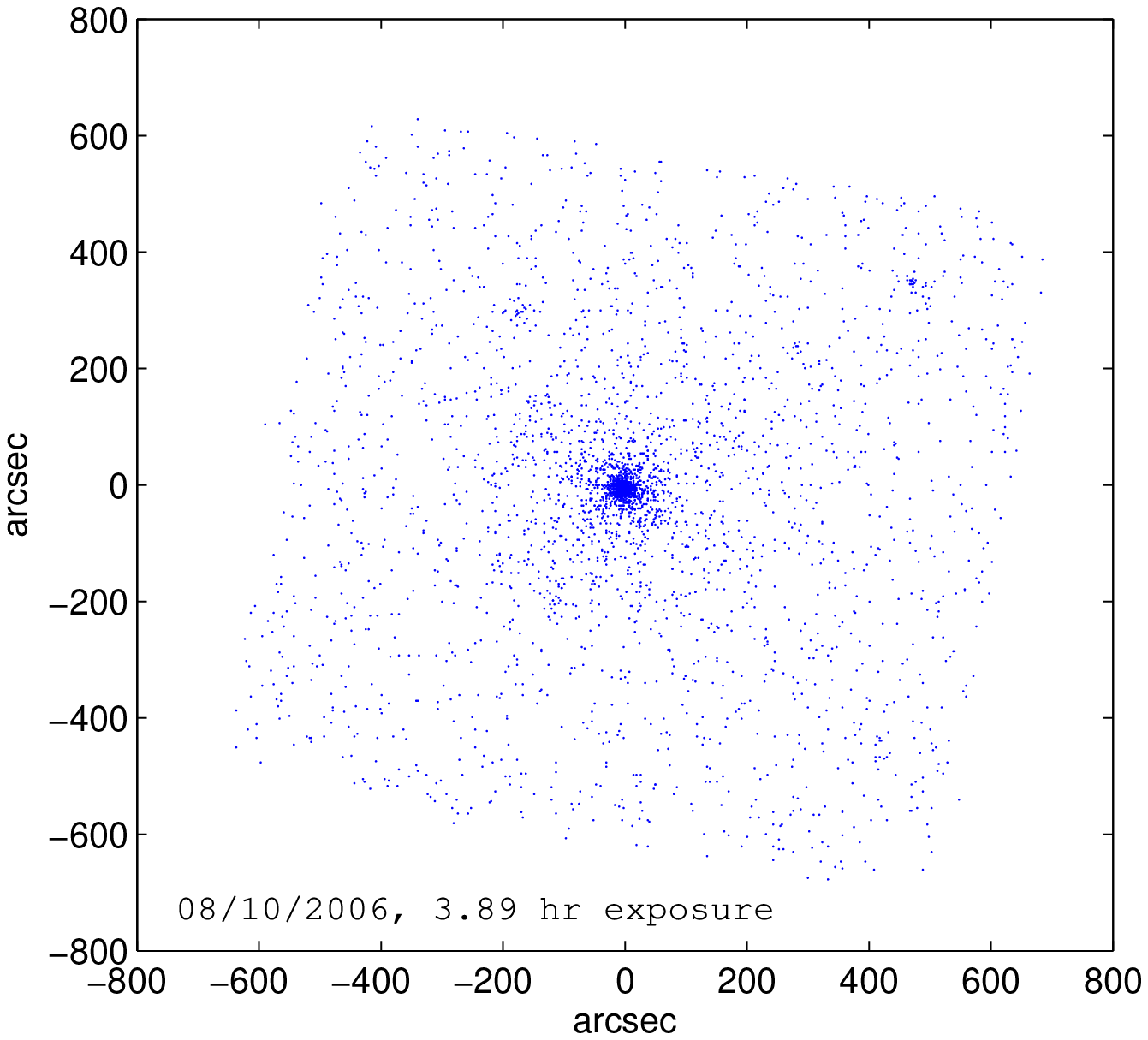}
\par

\caption{\label{fig:rings}The dust rings seen in the 1st (left, $\approx$31
hr after the burst) and 2nd (right, $\approx$90 hr after the burst)
${\it Swift}$/XRT observations, (energy range 2.5\textminus{}8.5
keV).}

\end{figure*}

SGR 1806\textminus{}20 intermediate burst of 2006 August 6 was first
reported by \citet{2006GCN..5416....1H,2006GCN..5419....1H}. It comprised
six separate bursts over $\approx$120 seconds, with the dominant
one lasting for $\approx$30 seconds and characterized by a measured
fluence of $2.4\times10^{-4}\mbox{\,\ erg}\,\mbox{cm}^{-2}$ and an
optically thin thermal bremsstrahlung (OTTB) spectrum of $kT=20$
keV within the Konus-Wind range of 20$-$200 keV \citep{2006GCN..5426....1G}.
Following this report, ${\it Swift}$/XRT took 12 observations in
Photon Count mode, starting 30 hours after the burst and ending 14
days later, corresponding to observation ID 00035315002\textminus{}00035315013.
For our analysis we used the four earliest observations (see Table
\ref{tab:observations}, for 1st and 2nd observations see Figure \ref{fig:rings}).
The other observations did not show a significant signal and were
only used implicitly for background consistency check. ${\it Swift}$/XRT
is sensitive to photons in the energy range 0.2\textminus{}10 keV,
and the dust scattered signal in the observations is only evident
in the range 2.5\textminus{}8.5 keV. Below this range the optical
depth is larger than unity and the assumption of a single scattering
does not hold. In order to allow energy dependent analysis without
diluting the signal too much we split each observation data into two
energy bands: 2.5\textminus{}4.5 and 4.5\textminus{}8.5 keV. We created
exposure maps for each observation and each energy band with the FTOOLS
task XRTEXPOMAP, to correct for the vignetting, the dead detector
areas, and the excluded regions. For each observation we extracted
a radial profile centered at the position of SGR 1806\textminus{}20
given by FTOOLS task XRTCENTROID. Radial bin width of 8 pixels (19$''$)
was chosen so it overlaps with average ${\it Swift}$/XRT point-spread
function (PSF) of $\cong20''$ (half power diameter, \citealp{2005SSRv..120..165B}).
We estimated the background photon count per energy band for each
observation by taking the average photon count at a radial distance
where no dust scattered signal is expected and adjusting for the area
of each radial bin and for its average exposure time derived from
the exposure map. The background thus measured agrees across all 12
observations with a photon count $C=1.40\pm0.13\times10^{-1}\,\mbox{ph\,\mbox{s\ensuremath{^{-1}}}}$
for the 2.5\textminus{}4.5 keV band and $C=9.57\pm1.20\times10^{-2}\mbox{\,\ ph\ensuremath{\,}\mbox{s\ensuremath{^{-1}}}}$
for the 4.5\textminus{}8.5 keV band over the XRT detection area.

The profile cannot be explained by the scattered signal and the
background alone, and is found to consist of a third component
\textminus{} a fading halo around the SGR. This halo is wider than
the central source PSF and therefore also contains one or more of
the following contributions: an integration over dust scattered
rings originated from fading post-burst source emission at times
between the burst and the observation; dust scattered rings
unresolved from the source due to dust clouds adjacent to the
source; and multiple scattering. Lacking knowledge about the
post-burst source luminosity as a function of time and confined by
the XRT resolution, we model this halo as a two parameters King
function $C\left(1+\left(\frac{r}{30}\right)^{2}\right)^{-\beta}$.
This component has no significant effect on our background
evaluations since the King function is negligible at the radial
distance used for background measurement.

\begin{table*}
\centering
\caption{\label{tab:observations}${\it Swift}$/XRT observations used in the
analysis. Times are given in hours, start and end times are relative
to the burst trigger. Total duration is the time elapsed between the
start and end time while exposure time is the net time the detector
was actively collecting signal.}

\hspace{1 pt}

\begin{tabular}{ccccc}
\hline
Obs. ID & Start time & End time & Total duration & Exposure time\\
 & {[}hr{]} & {[}hr{]} & {[}hr{]} & {[}hr{]}\\
\hline
00035315002 & 30.50 & 31.17 & 0.66 & 0.66\\
00035315003 & 82.22 & 104.81 & 22.59 & 3.89\\
00035315004 & 112.94 & 129.17 & 16.23 & 1.89\\
00035315005 & 141.93 & 153.29 & 11.36 & 1.88\\
\hline
\end{tabular}
\end{table*}

\section{Dust scattered x-ray rings}

An SGR flare can be considered a special case of a varying source,
where the flare is approximately a delta function in time, and the
result of an intervening dust screen is a ring expanding with time.
For small scattering angles and a single scattering screen, geometry
dictates \begin{equation}
\theta(t)\cong\left[\frac{2c}{d}\frac{\left(1-x\right)}{x}t\right]^{1/2}\label{eq:geometry}\end{equation}
 where $\theta(t)$ is the measured ring angle with respect to the
line of sight, $d$ is the source distance, $x$ is the screen's location
expressed as a fraction of $d$, and $t$ is the time passed since
the direct flare observation.

We fit the observations to a model based on
\citet{1998ApJ...503..831S}, with differential scattering cross
section as described in \citet{2010ApJ...710..797R}. The original
model calculates the halo observed by the scattering of x-rays from
a persistent source over a continuous distribution of intervening
dust. We are interested in the rings produced by the scattering of
an impulsive emission over one or more molecular clouds, namely a
discrete distribution of intervening dust. Our configuration is
simpler and it allows us to derive an analytic solution. We
therefore modify the original derivation by
\citet{1998ApJ...503..831S} to obtain an analytic expression for the
flux in a ring produced by the scattering of a flare on a single
thin dust screen, i.e. the scattered flux observed at an angular
distance $\theta_{{\rm obs}}(t)$ from the central source, as a
function of photon energy. The modification is straight forward and
we therefore do not repeat the derivation of
\citet{1998ApJ...503..831S} here, and just highlight the
modifications. Keeping the original notations, we replace the dust
distribution along the line of sight with the dust column density
across the scattering screen, $N_{\rm dust}$ and the dust grain size
distribution with $N_{\rm a}$, such that $N_{\rm dust}=\int N_{\rm
a}da$ where $a$ is the radius of the dust grain. We further replace
the flux reaching the scatterer, used in the original derivation, by
the flare's total fluence reaching it. In addition, the solid angle which the scattered observed signal occupies,
$d\Omega^{\prime}$, is $2\pi\theta(t) d\theta$ where $d\theta$ is
the change in the observed angle during the observation time $dt$.
Using Equation \ref{eq:geometry} we find
$d\Omega^{\prime}=g\left(x\right)dt$ where
$g\left(x\right)=\frac{2\pi c}{d}\frac{\left(1-x\right)}{x}$. The
observed scattered flux per photon energy $E$, $F_E$ (energy
observed in a ring per photons of energy $E$ per unit area per unit
time $dt$)  is thus:
\begin{equation}
F_E\left(\theta_{\rm obs}[t]\right)=S_E \frac{g(x)}{(1-x)^2}\int_{a_{{\rm min}}}^{a_{{\rm
max}}} N_{\rm a}\left(\frac{d\sigma}{d\Omega}\right)da
\label{eq:dwek}
\end{equation}
where $S_{E}$ is the flare's direct (vs. scattered) observed fluence
per photon energy $E$ and $d\sigma(E,a,\theta_{{\rm scat}})/d\Omega$
is the differential scattering cross section of an x-ray with energy
$E$ by a dust grain with radius $a$ at an angle $\theta_{{\rm
scat}}$.

As scattering angles are small, we substitute $\theta_{{\rm
scat}}=\theta_{{\rm obs}}/(1-x)$. For the dust grain size
distribution we assume a power-law $N_{\rm a}=Aa^{q}$ where $a_{{\rm
min}}\leq a\leq a_{{\rm max}}$ \citep{1977ApJ...217..425M}. For the
differential cross section we use the Gaussian approximation of the
Rayleigh-Gans theory, valid for small scattering angles and energies
above 1keV, following \citet{1957lssp.book.....V}:
\begin{equation}\label{eq sigma}
\frac{d\sigma(E,a,\theta_{{\rm scat}})}{d\Omega}\cong C_{{\rm
dust}}a^{6}\exp\left(-\frac{\theta_{{\rm
scat}}^{2}}{2\tilde{\theta}_{{\rm scat}}^{2}}\right)
\end{equation}
where $C_{{\rm dust}}$ depends on dust components atomic charge,
mass number, density, and scattering factor, while
$\tilde{\theta}_{{\rm scat}}$ is given by
\citet{1986ApJ...302..371M}
\begin{equation} \tilde{\theta}_{{\rm
scat}}\left(a,E\right)=10.4\frac{1}{(E/1\mbox{keV})(a/0.1\mu\mbox{m})}\,{\rm
arcmin}\label{eq:typical angle}
\end{equation}
Inserting the cross section and dust grain size distribution we
obtain
\begin{equation}
\begin{array}{l}
  F_{E}(\theta_{{\rm obs}},q,\hat{a}_{{\rm max}},\hat{a}_{{\rm
min}})= \\
\\
  B_{E}\int_{\hat{a}_{{\rm min}}}^{\hat{a}_{{\rm
max}}}\hat{a}^{q+6}\exp\left\{-\frac{1}{2}\left(\frac {\theta_{\rm
obs}}{10.4'}\right)^2\left(\frac{\hat{a}}{0.1\mu\mbox{m}}\right)^2\left(\frac{E}{1\mbox{keV}}\right)^2\right\}d\hat{a}\label{eq:model}
\end{array}
\end{equation}
where $\hat{a}\equiv\frac{a}{1-x}$, and
\begin{equation}
B_{E}=S_{E}C_{{\rm dust}}A\frac{2\pi
c}{d}(1-x)^{q+6}/x\label{eq:normalization}
\end{equation}

The differential cross section (Equation \ref{eq sigma}) implies
that a dust grain of size $a$ scatters effectively below a
characteristic angle $\tilde{\theta}\left(a\right)$, above which the
differential cross section decays rapidly. Thus, for a grain size
distribution more gradual than $a^{-7}$, regardless of its exact
form, the scattering at an angle $\theta_{\rm scat}$ is dominated by
the grain size matching Equation \ref{eq:typical angle},
\begin{equation}
a_{\rm scat}(\theta_{\rm scat},E)\approx \frac{0.1}{(E/1\mbox{keV})(\theta_{\rm scat}/10.4 {\rm arcmin})}\mu\mbox{m}
\end{equation}
The observed scattered flux can then be approximated as
$F_{E}(\theta)\propto N_{\rm a_{\,\rm scat}}a_{\rm scat}^7$. If the
grain size distribution is limited to the range $a_{{\rm min}}\leq
a\leq a_{{\rm max}}$ then at small angles (early time)  $a_{\,\rm
scat} > a_{{\rm max}}$ and the scattering is dominated by the
largest grains, namely it is constant in angle (time) and $F_{E}
\propto N_{a_{\rm max}}a_{{\rm max}}^7$. At large angles (late time)
 when $a_{\,\rm scat} < a_{{\rm min}}$ the scattering is dominated by the
Gaussian tail of the smallest grains cross-section, and the flux
falls exponentially. Therefore, defining
$\tilde{\theta}=\left(1-x\right)\tilde{\theta}_{{\rm scat}}$,
Equation \ref{eq:model} can be approximated as
\begin{equation}
F_{{\rm E}}(\theta)\propto\left\{ \begin{array}{ll}
const & \theta_{{\rm obs}}\ll\tilde{\theta}\left(a_{{\rm max}}\right)\\
\theta_{{\rm obs}}^{-(q+7)} & \tilde{\theta}\left(a_{{\rm max}}\right)\ll\theta_{{\rm obs}}\ll\tilde{\theta}\left(a_{{\rm min}}\right)\\
\exp\left[-\frac{\theta_{{\rm
obs}}^{2}}{\tilde{\theta}^{2}\left(a_{{\rm min}}\right)}\right] &
\theta_{{\rm obs}}\gg\tilde{\theta}\left(a_{{\rm
min}}\right)\end{array}\right.\label{eq:approx1}\end{equation} It is
useful to express this approximation in terms of the flux dependence
on time, using Equation \ref{eq:geometry}:
\begin{equation}
F_{{\rm E}}(t)\propto\left\{ \begin{array}{ll}
const & t\ll\tilde{t}\left(a_{{\rm max}}\right)\\
t^{-(q+7)/2} & \tilde{t}\left(a_{{\rm max}}\right)\ll t\ll\tilde{t}\left(a_{{\rm min}}\right)\\
\exp\left[-\frac{t}{\tilde{t}\left(a_{{\rm min}}\right)}\right] &
t\gg\tilde{t}\left(a_{{\rm
min}}\right)\end{array}\right.\label{eq:approx2}\end{equation} where
$\tilde{t}\left(a,E\right)=\frac{d\,
x}{2c\left(1-x\right)}\tilde{\theta}^{2}\left(a,E\right)$. These
expressions describe three flux regimes \textminus{} (i) a constant
set by scattering over the largest grains. (ii) A power-law decay
where the size of the grains that dominate the scattering vary with
time and (iii) an exponential decay set by the scattering cross
section of the smallest grains. Therefore, observing the evolution
of $F_E$ with time is a direct probe of the dust grain size
distribution. $a_{{\rm max}}$ is probed by observations covering the
time of transition between the constant and the power-law regime,
while the index $q$ is obtained by power-law regime observations.
Probing $a_{{\rm min}}$ requires a burst bright enough for the
scattered signal to overcome the background at the low end of the
power-law regime.

We note that Equation 10 in \citet{2003ApJ...598.1026D}, which is
an analytical approximation for the differential scattering cross
section describing his model of x-ray scattering by dust, can be closely
matched by the above scattering model by choosing $a_{{\rm max}}=0.33\,\mu\mbox{m}$
and $q=-3$, which are typical values for ISM dust.

\section{Analysis \& Results}

To obtain our two objectives \textminus{} constraining the distance
to SGR 1806\textminus{}20 and the properties of the dust in the
intervening molecular clouds, we take two different approaches in
analyzing the XRT observations. First (Section
\ref{sub:Analysis-clouds}), we assume positions and visual
extinctions of molecular clouds along the line of sight based on
CE04 in order to get an estimate of the distance to SGR
1806\textminus{}20. We test the dependence of this analysis on a
dust scattering model and find it to be negligible. Therefore our
conclusion is based on the accuracy of CE04 clouds distribution, and
is mainly limited by the XRT resolution and sensitivity.

In the second (Section \ref{sub:Analysis-dust-prop}) we relax our
assumptions regarding the distance to the scattering clouds, and motivated
by observations we assume a single predominant scattering screen.
We keep the location of this screen a free parameter, and use Equation
\ref{eq:model} to put simultaneous constraints on both the dust screen
location and the dust properties, using a full data set including
both spectral and temporal flux evolution. The results of this analysis
are limited mainly by the delay of the first observation and by the
sensitivity of the XRT detector.

\subsection{Distance estimate based on known molecular clouds along the line
of sight\label{sub:Analysis-clouds}}

Equation \ref{eq:dwek} expresses the observed scattered flux
as a function of: (1) The scatterer's location relative to the
source (i.e., $x$) and (2) Properties of the scattering dust.
Therefore, if the location of the scatterers and the properties of
the dust are known, one can extract the distance to the source. In
fact, as we show here, it is enough to know the location of the dust
screens along the line of sight and their dust column densities to put
strong constraints on the source distance, even if the grain size
distribution, $N_a$ is not well known. CE04 provide the locations of
the scattering clouds along the line of sight to SGR
1806\textminus{}20, as well as their optical extinction $A_{V}$,
which indicate on their dust column densities. Using this data we
calculate the scattered flux radial profile for different locations
of the source. We then compare these {}``synthetic''
profiles with the observed profiles and use the best fit profile to
infer the distance.

Table \ref{tab:clouds} lists properties of the molecular
clouds along the line of sight, adapted from Table 1 of CE04.
Following radial velocity measurements, each cloud has two distance
solutions and was therefore attributed with a near distance and a
far distance, as well as optical extinction $A_{V}$. For eight
clouds CE04 ruled out one of these distances due to additional
considerations. We assume that the $A_{V}$ given for each cloud is
proportional to its share of the clouds' total dust column density
(i.e. $A_{V}\propto A$ in Equation \ref{eq:normalization}). Equipped
with the clouds' positions and dust shares, we assume a single dust
grain size distribution across all clouds and set $a_{{\rm
max}}=0.3\,\mu\mbox{m}$, $a_{{\rm min}}=0\,\mu\mbox{m}$, $q=-3$.
Assuming a source distance we use Equations \ref{eq:geometry} and
\ref{eq:model} to calculate the scattered flux from each of the
intervening clouds and construct a synthetic flux radial profile at
any given time and energy band. In order to compare the profile
calculated using Equations \ref{eq:geometry} and \ref{eq:model} to
the observations, it must be smeared by the PSF, the exposure
duration and the width of the scattering clouds. We therefore
introduce the ${\it Swift}$/XRT PSF function
\citep{2005SPIE.5898..348M}, and we account for the observations
actual exposure durations to widen the profile beyond the PSF
effect. Assuming a typical molecular cloud width of $\sim100\,{\rm
pc}$, the consequent smear is negligible. We repeat this calculation
for an SGR 1806\textminus{}20 distance that vary in the range
5\textminus{}25 kpc with a 0.1 kpc resolution. Note that the
synthetic profiles are calculated up to a normalization factor per
energy band, accounting for the unknown direct fluence from the
source $S_{{\rm E}}$, as well as both $C_{\rm dust}$ and the
constant of proportionality in $A \propto A_{V}$.

\begin{table}
\centering
\caption{\label{tab:clouds}Molecular clouds along the line of sight to LBV
1806\textminus{}20, taken from CE04. The bottom three clouds, lacking
unambiguous distance determinations, are quoted with both near/far
distances ($1\sigma$ errors). CE04 do not determine whether the three
last clouds in this table are in front or behind the Galactic center,
allowing two distances to each cloud. In the text we refer to all
three clouds being in front of the Galactic center as the `near' configuration,
while all three clouds being at their far distance is the `far' configuration.}

\hspace{1 pt}

\begin{tabular}{lcc}
\hline
Name & $A_{V}$ & Distance\tabularnewline
 & {[}magnitude{]} & {[}kpc{]}\tabularnewline
\hline
MC-16 & 8.6$\pm$1.7 & 4.5$\pm$0.3\tabularnewline
MC4 & 3.9$\pm$0.8 & 0.2$\pm0.5$\tabularnewline
MC13A & 10.0$\pm$2.0 & 15.1$\pm$0.9\tabularnewline
MC13B & 3.0$\pm$0.6 & 4.5$\pm$0.3\tabularnewline
MC24 & 5.6$\pm$1.1 & 3.0$\pm$0.5\tabularnewline
MC30 & 0.8$\pm$0.2 & 3.5$\pm$0.35\tabularnewline
MC38 & 1.1$\pm$0.2 & 4.2$\pm$0.3\tabularnewline
MC44 & 1.6$\pm$0.3 & 4.5$\pm$0.3\tabularnewline
MC73 & 6.0$\pm$1.2 & 5.7/11.0$\pm$0.15\tabularnewline
MC87 & 1.4$\pm$0.3 & 6.1/10.6$\pm$0.15\tabularnewline
MC94 & 0.5$\pm$0.1 & 6.2/10.5$\pm$0.15\tabularnewline
\hline
\end{tabular}
\end{table}

The actual observed signal contains two additional features beyond
the dust scattered flux, and we therefore model the observed radial
flux profile as a sum of the three following contributions: (a) the
synthetic profile of the dust scattered signal, calculated as
described above (b) the measured background profile, and (c) the
halo around the SGR described in Section \ref{sec:Observations}.
Therefore, our model contains the following free parameters: the
halo amplitude and power-law per observed profile, the normalization
factor of the scattered signal per energy band, and the SGR
distance.

We compare this model with the observed profiles from the first two
epochs in Table \ref{tab:observations}, where the observed dust
scattered ring is most evident, with each epoch split into two
energy bands. For each distance we use a $\chi^{2}$ fit to find the
halo amplitude and power-law per observed profile, and the
normalization factor per energy band. We then calculate $\chi^{2}$
as a function of distance and choose the distance yielding the
overall minimal $\chi^{2}$ as the best fit distance.

Three of the clouds lack an unambiguous distance determination and
have instead both a near and a far distance estimates, similar for
all three clouds. To account for this uncertainty we prepare two
sets of scattered signal profiles as a function of SGR distance, one
with all three clouds set at their 'near' position, the other with
all three at their 'far' position, and repeat the above procedure
for both sets. We verify that other combinations, where some
of the clouds are at an opposite location, result in intermediate
SGR locations and therefore the 'far' and 'near' configurations
bracket the constraints on the location of SGR 1806\textminus{}20.

\begin{figure}
\begin{centering}
\includegraphics[scale=0.25]{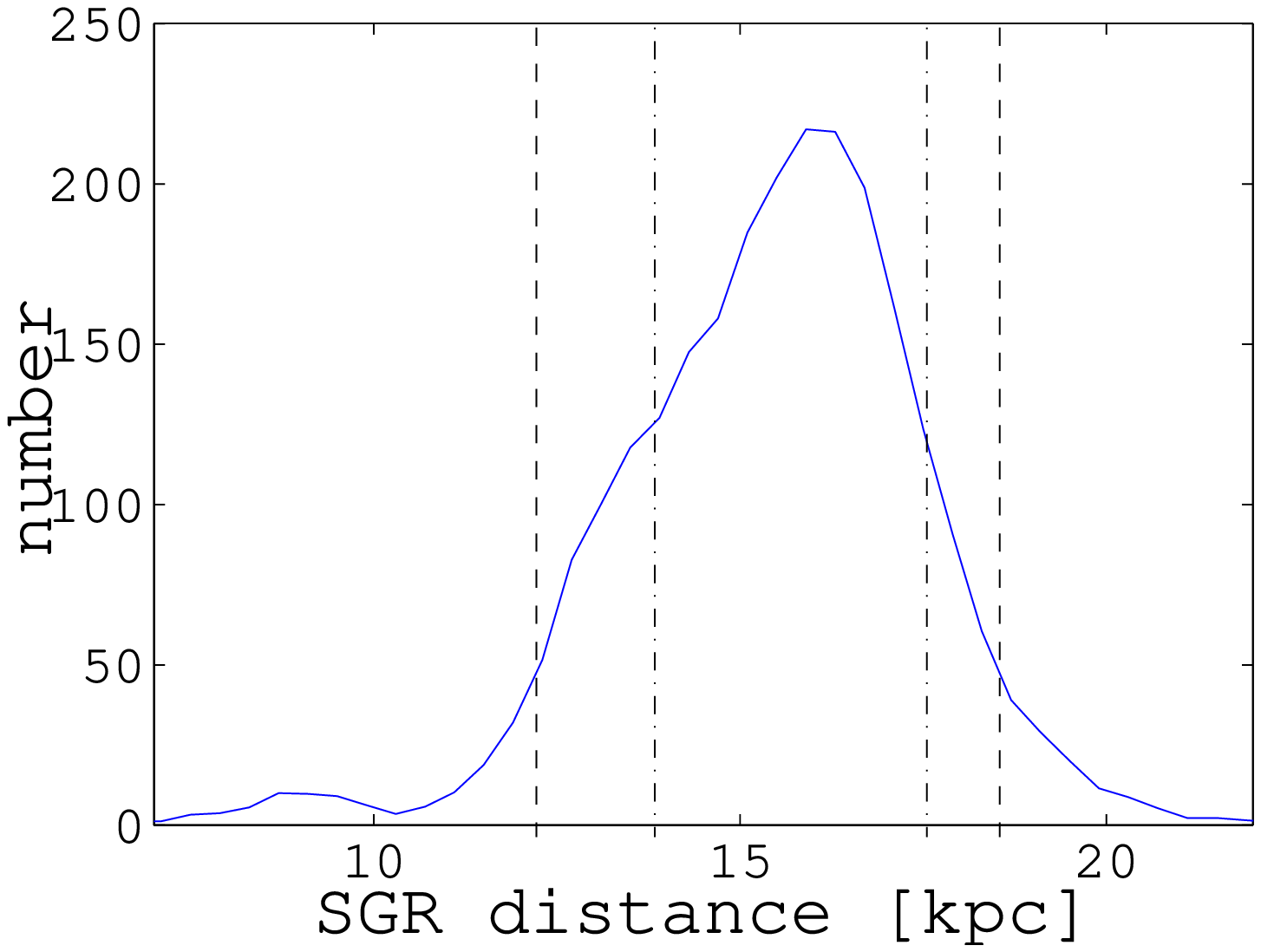}
\includegraphics[scale=0.25]{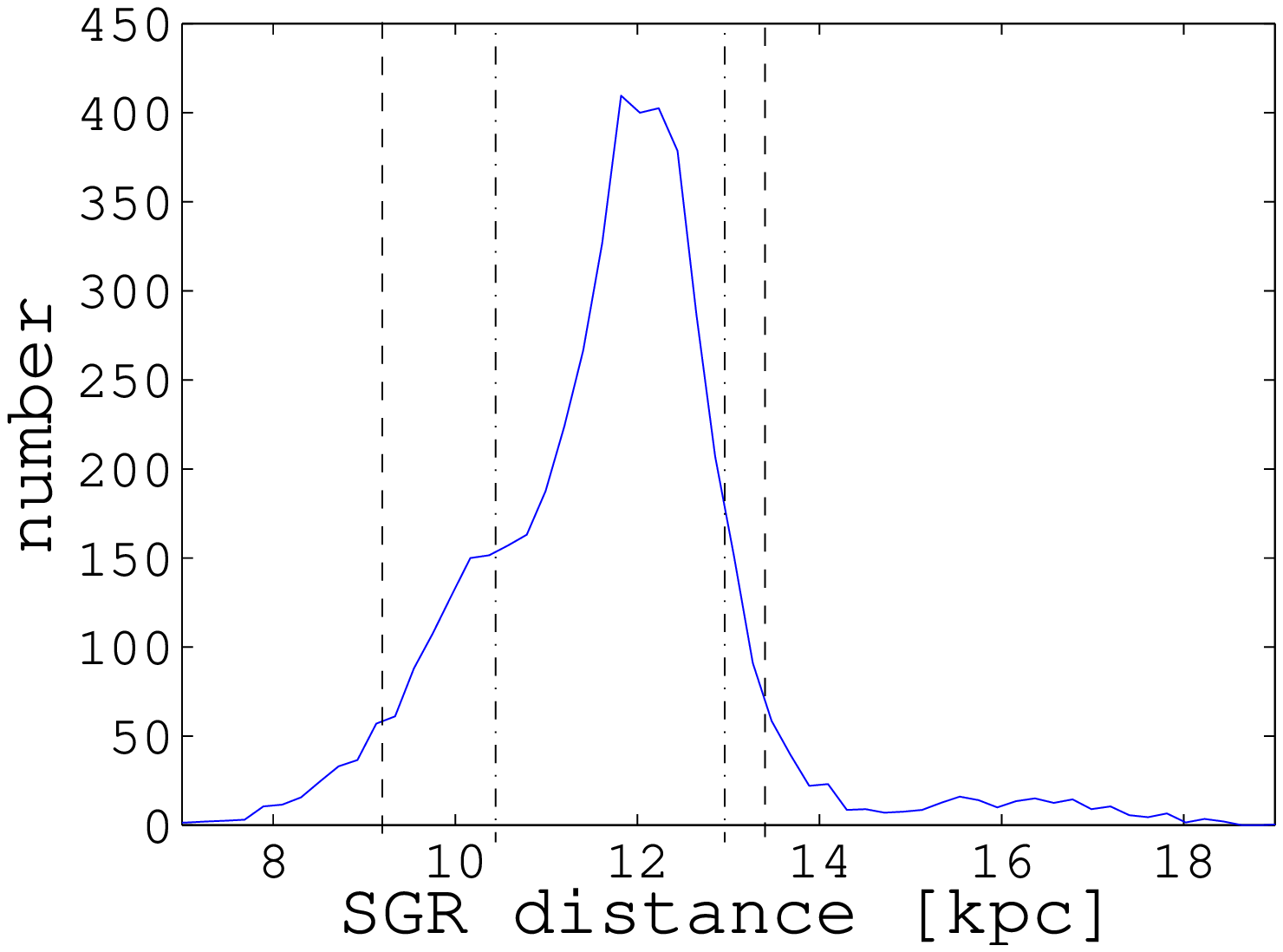}
\par\end{centering}

\caption{\label{fig:hist}The distributions of the distance to SGR 1806\textminus{}20
obtained by Monte Carlo simulations, with 68\% and 90\% confidence
limits marked. The left (right) distribution matches the 'near' ('far')
clouds configuration, where the three clouds with ambiguous locations
are fixed at their near (far) locations (see Table \ref{tab:clouds}).
The best fit distance is $16{}_{-3.8}^{+2.6}$ kpc for the 'near'
configuration and $11.9_{-2.5}^{+1.6}$ kpc for the 'far' configuration
(90\% confidence).}

\end{figure}

In order to estimate the distance error we use Monte Carlo simulations
(following the prescription in \citealt{1986nras.book.....P}). For
each Monte Carlo realization we generate a pseudo binned radial profile
by drawing the {}``observed'' radial bins photon counts from a Poisson
distribution around the best fit radial profile. We further prepare
synthetic scattered profiles per SGR distance by drawing clouds' positions%
\footnote{CE04 use a systematic error of $\pm10\mbox{km\,\ s}^{-1}$ on their
measured velocity of the clouds in order to derive $>2\sigma$ distance
errors. We therefore adopt a $\pm5\mbox{km\,\ s}^{-1}$ $1\sigma$
velocity error for estimating the distance error to all the clouds
using the Galactic rotation curve toward the line of sight presented
in Figure 5 of \citet{2004ApJ...610L.109F}, based on \citet{1993A&A...275...67B}.
The resulting distance errors are presented in Table \ref{tab:clouds}.%
} and optical extinctions (from Table \ref{tab:clouds}) assuming an
approximately normal distribution for both. We repeat the fitting
procedure described above for each of 10,000 Monte Carlo realizations
and thus obtain the probability distribution of the SGR distance (see
Figure \ref{fig:hist}).

To account for the dependence of the distance on the dust grain size distribution
we run the complete procedure described above for nine different parameter
choices of Equation \ref{eq:model}, with the permutations of $a_{{\rm max}}=0.15,0.3,0.6\,\mu\mbox{m}$
and $q=-3.5,-3,-2.5$. We find the best fit distance scatter due to
the model choice to be smaller than the scatter due to the errors
of the binned signal and of the clouds' positions and opacities (a
scatter of $\sim1$ kpc between the two extreme model parameter choices,
compared to $\sim3.5$ kpc $\left(1\sigma\right)$ scatter due to
the errors). We therefore set the intermediate values of $a_{{\rm max}}=0.3\,\mu\mbox{m}$
and $q=-3$ as our model parameters for the analysis. In Section \ref{sub:Analysis-dust-prop} we measure $q$ and find a lower limit for $a_{\rm max}$, which are consistent with the parameters we choose for this analysis, and which errors are consistent with the range we use for this parameter independency check. The absence of a measured upper limit for $a_{\rm max}$ has no influence on this analysis since the dominant clouds fall within the power-law regime (see Equation \ref{eq:typical angle}) for any $a_{\rm max}$ bigger than the value chosen here.

\begin{figure*}
\begin{centering}
\includegraphics[scale=0.5]{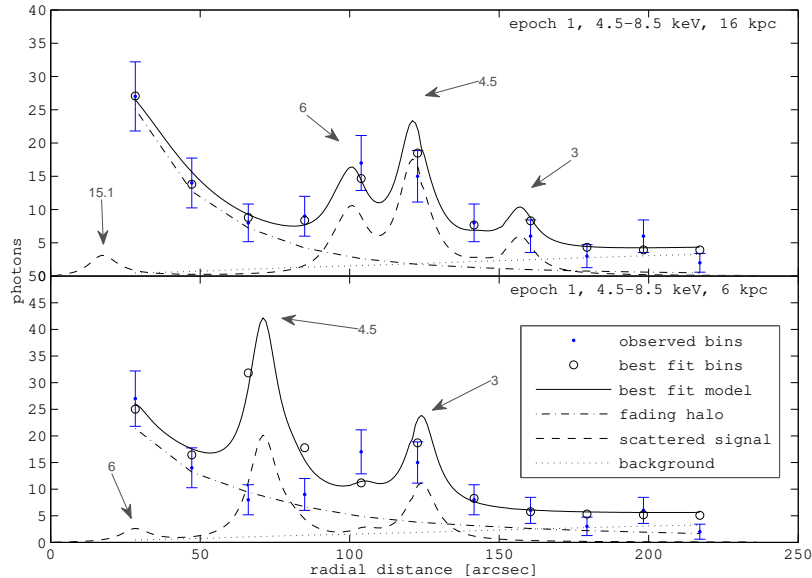}
\par\end{centering}

\caption{\label{fig:clouds}A comparison between the best fit $\chi^{2}$ obtained
for SGR 1806\textminus{}20 located at 6 kpc vs. 16 kpc, with clouds
at the 'near' clouds configuration (see Table \ref{tab:clouds}).
Circles mark the best fit model binnned photon counts while dots with
error bars (Poisson, $1\sigma)$ mark the measured bin photon counts.
Arrows indicate the positions in kpc of the main clouds concentrations
corresponding to each peak. An SGR distance of 16$\,$kpc evidently
provides a good fit to the data, while a distance of 6$\,$kpc is
ruled out at a $>3\sigma$ level. Model components: Dotted line is
background, dotted-dashed is the fading halo, dashed is the calculated
signal profile, continuous line is sum of the three components of
the best fit model.}

\end{figure*}

The effect of the SGR position on the scattered signal and on the
fit quality is demonstrated in Figure \ref{fig:clouds}. For an SGR
at 16$\,$kpc, the clouds concentrations at 4.5 and 6$\,$kpc nearly
merge, overlapping with the measured signal, while the 3$\,$kpc clouds
are negligible. In contrast, with the SGR at 6$\,$kpc the clouds
concentrations are more separated, with the dominant 4.5$\,$kpc cloud and
the measured signal largely unsynchronized. For the 'near' configuration,
the best fit distance is $16{}_{-3.8}^{+2.6}$ kpc (90\% confidence),
with $\chi^{2}/{\rm dof}=43.3/47$. For the 'far' configuration, best
fit distance is $11.9_{-2.5}^{+1.6}$ kpc (90\% confidence) with $\chi^{2}/{\rm dof}=45.9/47$.
Since the goodness of fit for both configurations is similar we use
their combination to set conservative limits, with a lower limit of
$9.4$ kpc and an upper limits of $18.6$ kpc on the distance to SGR
1806\textminus{}20 at a 90\% confidence level.

We note that taking the main bulk of the dust along the line of sight
according to CE04, concentrated at 4.5$\,$kpc, and substituting in
Equation \ref{eq:geometry} with our observed ring's $\theta$ and
$t$, results in a distance of 15$\,$kpc for SGR 1806\textminus{}20.
Our above analysis wraps this crude estimation with the confidence
limits permitted by the XRT observations and the CE04 errors.

The weakest link of our analysis is the accuracy of the clouds' locations
as reported in CE04. As an example, \citet{2004ApJ...610L.109F} suggested,
based on a different method for measuring the radial velocity, a distance
of 11.8 kpc (vs. 15.1$\pm$0.9 of CE04) to cloud MC13A. Although this
cloud has minor influence on our analysis, the discrepancy in this
measurement demonstrates the limitation of our method. However, the
goodness of fit we obtain, $\chi^{2}/\mbox{dof}\approx1$, provides
a consistency check for the CE04 clouds distribution along the line
of sight. In Section \ref{sec:Future-SGRs-observations} we argue
that future post-burst observations could resolve the location of
a few individual clouds. This would potentially permit an SGR distance
estimate that depends, independently, on the location of each of the
resolved clouds, thereby resulting in a more robust and accurate distance
measurement for this SGR.

\subsection{Constraints on dust properties\label{sub:Analysis-dust-prop}}

Here we assume a single predominant dust screen at an unknown
a-priori location, thus dropping our previous rely on CE04 results.
This assumption is motivated by the fact that the XRT observations
show a single ring. Fitting the observed radial profile fluxes and
photon energies using Equation \ref{eq:model}, we obtain best fit
values for $\hat{a}_{{\rm max}}$, $q$ and $B(E)$. Since
$\hat{a}_{{\rm max}}=a_{{\rm max}}/\left(1-x\right)$ the fit results
in a joint constraint on $a_{{\rm max}}$ and $x$. The model we use
assumes a power-law grain size distribution $N_{\rm a}\propto
a^{q}$ with grain size $a_{{\rm min}}<a<a_{{\rm max}}$, leading to
the three scattered flux regimes described in Equation
\ref{eq:approx1}. Thus, constraining $q$ requires observations at
epochs or angles within the power-law regime, while constraining
$\hat{a}_{{\rm max}}$ requires the further inclusion of earlier
observations, towards the constant flux asymptote.

Fitting the observations to the dust model is done in two
steps. First we find the integrated scattered signal flux within the ring and its associated flux probability distribution (from simulations) per epoch (for the four epoches in Table
\ref{tab:observations}) per energy band by fitting the observed data
to a model. Then we use these fluxes as a function of epoch and
energy, and their probability distributions, to obtain best fit
values for $\hat{a}_{{\rm max}}$, $q$ and $B(E)$ by means of a
Maximum likelihood fit using Equation \ref{eq:model}.

\begin{figure}
\includegraphics[scale=0.45]{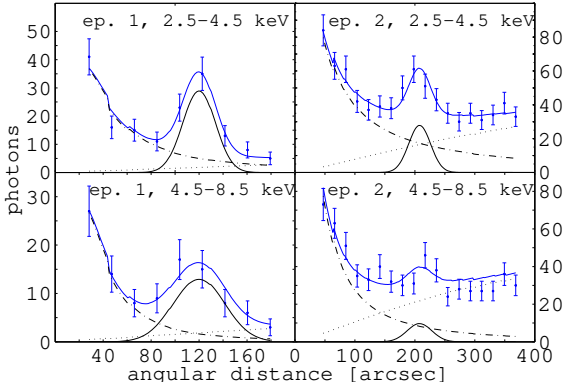}
\caption{\label{fig:fluxes}$\chi^{2}$ fits for the first two epochs
in the two energy bands. The dotted line is background,
dotted-dashed is the fading halo, bottom continuous line is the
scattered signal Gaussian and top line is the sum of the three
components. Dots with error bars (Poisson, $1\sigma$) are the binned
measured photon counts. The best fit Gaussian area translates to the
flux per epoch and energy band (Table \ref{tab:Best-fit-fluxes}),
flux errors distribution are found using Monte Carlo simulations
(see text).}
\end{figure}

We model the dust scattered signal as a Gaussian profile,
thus accounting for ${\it Swift}$ PSF and exposure duration effect,
assuming that the effect of the clouds' width is negligible. In
order to calculate the scattered signal net flux per energy band per
epoch, we fit each observed radial profile to a model composed of
the following three components: (a) the Gaussian profile
representing the scattered signal (b) the measured background profile,
and (c) the halo around the SGR described in Section \ref{sec:Observations}. The
contribution of each component to each bin is weighted according to
the bin's average exposure time due to the exposure maps. We run a
minimum $\chi^{2}$ fit with the fading halo amplitude and power, and
the Gaussian amplitude and width as the free parameters. For the
first observation we also
fit for the radius of the Gaussian center and obtain%
\footnote{An upper limit of 6.37 kpc for the dominant dust screen, regardless
of the source position, is obtained by substituting the observed ring's
$\theta$ and $t$ in Equation \ref{eq:geometry} (a 6.6 kpc upper
limit was reported by \citealt{2006GCN..5438....1G}).%
} $\theta=119\pm 5''\,(2\sigma)$. We fix the ring location of the other observations
using $\theta\left(t\right)\propto\sqrt{t}$ (Equation
\ref{eq:geometry}). Figure \ref{fig:fluxes} shows the best fit model
for the first two epochs. The best fit Gaussian area translates to
the photon count per second related to the scattered signal, which
we convert to flux using NASA's HEASARC web based tool
WebPIMMS\footnote{Using a power-law source model with photon index
-1, i.e. a blackbody Rayleigh-Jeans law
\citep[e.g.][]{2004ApJ...612..408F}, and $N_H=6\times 10^{22}\,{\rm
cm}^{-2}$ \citep{1994ApJ...436L..23S}.}.

We find probability distributions of each ring's integrated flux by
Monte Carlo simulations. The best fit radial profile from the $\chi^{2}$
fit is used as the base data set for preparing (assuming Poisson distribution
for the bins photon count) 5000 binned profile realizations. Collecting
the best fit flux of each realization we compile the flux probability
distribution.

\begin{figure}
\begin{centering}
\includegraphics[scale=0.5]{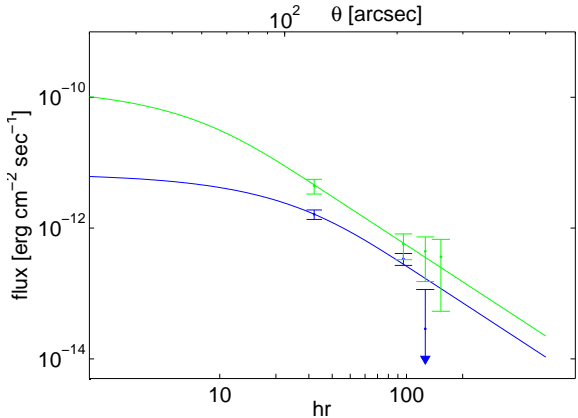}
\par\end{centering}

\caption{\label{fig:ML}Maximum likelihood best fit model (Equation \ref{eq:model}).
Lower line and accompanying dots and error bars (Table \ref{tab:Best-fit-fluxes})
are the 2.5\textminus{}4.5 keV band. The best fit result for the power-law
index $q$ is $-3.3_{-0.7}^{+0.6}$ ($1\sigma$), with errors determined
from Monte Carlo simulations. The best fit result for $\hat{a}_{{\rm max}}=a_{{\rm max}}/(1-x)$
is $0.33\,\mu\mbox{m}$. Since the first observation is almost completely
within the flux power-law regime (see Equation \ref{eq:approx1}),
we can not set an upper limit for $\hat{a}_{{\rm max}}$. The $2\sigma$
lower limit is $0.21\,\mu\mbox{m}$ (see Figure \ref{fig:contours}).}

\end{figure}

These distributions are then used as the input for a Maximum
Likelihood fit, where we find the best fit values for $\hat{a}_{{\rm
max}}$ and $q$, along with two normalization factors $B_{E}$, one
per energy band. For a given set of parameters
$\hat{a}_{{\rm max}},q$ and $B(E)$ we use Equation
\ref{eq:model} to calculate the flux at each observation epoch and
energy band. We then evaluate the likelihood of this set using the
measured flux distribution. The set of parameter values which yields
the highest product of likelihoods for all epochs and energy bands
is the best fit. Figure \ref{fig:ML} shows the $1\sigma$ bars of
the flux distributions, and the corresponding best fit model.

Estimation of the confidence limits of these parameters is
calculated by Monte Carlo simulations. For each Monte Carlo run we
generate a realization of the fluxes in Table
\ref{tab:Best-fit-fluxes}, with the ring's integrated flux values
drawn from the actual best fit distribution in Table
\ref{tab:Best-fit-fluxes}. The errors around the drawn flux values
are taken to be the same as the original errors in Table
\ref{tab:Best-fit-fluxes}. We fit 5000 Monte Carlo realizations for
$\hat{a}_{{\rm max}},q$ and $B(E)$, thus obtaining the probability
distribution of these parameters.

\begin{table*}
\centering
\caption{\label{tab:Best-fit-fluxes}Best fit ring's integrated fluxes used
for the maximum likelihood fit}
\hspace{1 pt}

\begin{tabular}{ccccc}
\hline
Epoch & Time since burst & Energy band & Flux & $\chi^{2}$/dof\tabularnewline
 & {[}hr{]} & {[}keV{]} & {[}$\mbox{erg}\,\mbox{cm}^{-2}\mbox{s}^{-1}${]} & \tabularnewline
\hline
1 & 30.8 & 2.5\textminus{}4.5 & $1.63\pm0.27\times10^{-12}$ & 2.95/4\tabularnewline
1 & 30.8 & 4.5\textminus{}8.5 & $4.45\pm1.13\times10^{-12}$ & 2.00/5\tabularnewline
2 & 93.5 & 2.5\textminus{}4.5 & $3.39\pm0.71\times10^{-13}$ & 6.15/15\tabularnewline
2 & 93.5 & 4.5\textminus{}8.5 & $5.73\pm2.45\times10^{-13}$ & 17.25/15\tabularnewline
3 & 121.1 & 2.5\textminus{}4.5 & $2.88\pm8.70\times10^{-14}$ & 7.59/11\tabularnewline
3 & 121.1 & 4.5\textminus{}8.5 & $4.43\pm2.92\times10^{-13}$ & 13.31/11\tabularnewline
4 & 147.6 & 4.5\textminus{}8.5 & $3.64\pm3.12\times10^{-13}$ & 16.83/11\tabularnewline
\hline
\end{tabular}
\end{table*}

\begin{figure}
\begin{centering}
\includegraphics[scale=0.5]{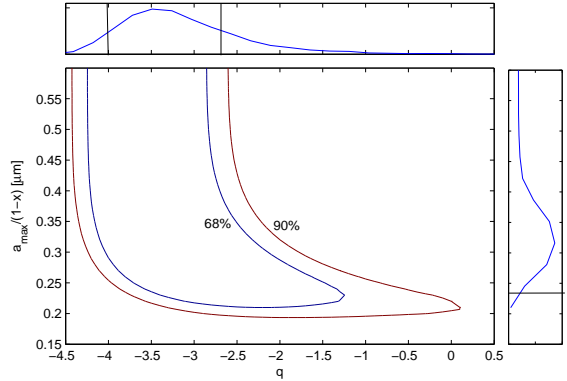}
\par\end{centering}

\caption{\label{fig:contours}68\% and 90\% likelihood contours of $q$ and
$\hat{a}_{{\rm max}}$ as determined from the Monte Carlo simulations.
Note the lack of an upper limit for $\hat{a}_{{\rm max}}$. The marginalized
distribution of each parameter are shown above ($q$) and to the right
($\hat{a}_{{\rm max}}$). The lines mark the 68\% confidence limits
for each single parameter, based on an integration of the probabilities
for $\hat{a}_{{\rm max}}\le3\mu\mbox{m}$.}

\end{figure}

Fluxes used for the maximum likelihood fit, with $1\sigma$ errors
and $\chi^{2}$ values, are listed in Table 1. The best fit result
for $q$ is $-3.3_{-0.7}^{+0.6}$ ($1\sigma$), consistent with the
commonly cited $-3.5$ from \citet{1977ApJ...217..425M} and with
an implicit $-3$ of \citet{2003ApJ...598.1026D}. The best fit result
for $\hat{a}_{{\rm max}}$ is $0.33\,\mu\mbox{m}$. As indicated by
the likelihood contours of $q$ and $\hat{a}_{{\rm max}}$ shown in
Figure \ref{fig:contours}, we are unable to constrain $\hat{a}_{{\rm max}}$
from above, which is the result of the earliest observation being
too adjacent to the power-law asymptote regime (see Figure \ref{fig:ML}).
We thus find only a lower limit of $0.21\,\mu\mbox{m}$ ($2\sigma$)
for $\hat{a}_{{\rm max}}$.

In our fit we constrain $\hat{a}_{{\rm max}}$, which is a
combination of $a_{{\rm max}}$ and $x$ ($\hat{a}_{{\rm
max}}\equiv\frac{a_{{\rm max}}}{1-x}$). In order to learn further
about either $x$ or $a_{{\rm max}}$ one needs to have prior
knowledge of either $a_{{\rm max}}$ or $x$, respectively. Since
$\theta\left(t\right)$ is known, a measurement of $x$ will provide
an SGR distance estimate that is independent of our previous
analysis. However, prior constraints on the distance to SGR
1806\textminus{}20, 6\textminus{}15 kpc (Section
\ref{sec:Introduction}), translate, together with the locations of
the ring, to $0.30<x<0.51$ (Equation \ref{eq:geometry}). Therefore,
improving these constraints on $x$ requires constraints on both
$\hat{a}_{{\rm max}}$ and $a_{{\rm max}}$ that are better than 25\%.
The uncertainty in the value of $\hat{a}_{{\rm max}}$ in the diffuse
ISM is larger, while its value in molecular clouds is practically
unknown. Therefore, our measured $\hat{a}_{{\rm max}}$, and even an
ideal errorless $\hat{a}_{{\rm max}}$, are useless for improving the
constraints on the distance to SGR 1806\textminus{}20. This
conclusion is similar to that of \cite{2010ApJ...710..227T}, which
find that the distance to the X-ray pulsar 1E1547.0\textendash{}5408
cannot be conclusively determined due to the uncertainty in the dust
properties, although they obtained superb data.

In the case of SGR 1806-20, however, there are robust (although
rather loose) limits on its distance. Thus, the prior constraints on
$x$ can be used to efficiently constrain $a_{{\rm max}}$, which is
of special interest in our case of molecular clouds. The robust
constraint of $d>6$ kpc for the SGR's distance
\citep{2005Natur.434.1112C} matches to $x<0.51$, which translates to
$a_{{\rm max}}>0.1\,\mu\mbox{m}$ ($2\sigma$).

The best fit results for the normalization factors are
$B(2.5-4.5{\rm keV})=2.89_{-0.81}^{+0.86}\times10^{-13}$
$\mbox{erg}\,\mbox{cm}^{-2}\,\mbox{s}^{-1}\,\mbox{asec}^{-2}$,
$B(4.5-8.5{\rm keV})=6.10_{-1.21}^{+1.27}\times10^{-12}$
$\mbox{erg}\,\mbox{cm}^{-2}\,\mbox{s}^{-1}\,\mbox{asec}^{-2}$. As a
consistency check, we use the ratio between the two normalization
constants to find the Hydrogen column density towards SGR
1806\textminus{}20. The normalization constant depends on photon
energy through $S_{E}$, the direct observed fluence. Given the
source spectrum and the x-ray extinction cross section per H nucleon
per energy $\sigma(E)$, $N_{H}$ can be extracted using
$S_{E}\propto\exp\{-\sigma(E)N_{H}\}$. Assuming the x-ray extinction
is mainly dust driven, we take the x-ray extinction cross section
per H nucleon due to dust for each energy band from
\citet{2003ApJ...598.1026D} (assuming $R_{V}=3.1$). For the burst
x-ray spectrum at our energy bands we follow
\citet{2004ApJ...612..408F} and \citet{2003AIPC..662...82O} in
adopting a double black body spectrum instead of the OTTB spectrum
commonly used to describe SGR bursts in the range $20-200\,{\rm
keV}$ (e.g. \citealt{2005GCN..4312....1G}). Using the average double
black body spectrum of $kT_{1}=3.4,\, kT_{2}=9.3,\,
L_{1}/L_{2}=0.85$ from \citet{2004ApJ...612..408F} with our best fit
ratio we get $N_{H}=9.64_{-4.0}^{+2.8}\times10^{22}\mbox{cm}^{-2}$,
consistent with \cite{1994ApJ...436L..23S}.

Reversing the last procedure we find the normalizations ratio between
the 2.5\textminus{}4.5 keV band and the 0.5\textminus{}2.5 keV band,
which we did not use due to lack of signal, to be 1000:1, explaining
the lack of signal beyond the background noise in the 0.5\textminus{}2.5
keV energy band.

\section{Future SGRs observations}\label{sec:Future-SGRs-observations}

The accuracy of the derived distance to SGR 1806\textminus{}20 and
the properties of the dust along the line of sight are restricted
by large error bars due to the weak signal and low resolution of the
${\it Swift}$/XRT, and by the long delay between the burst and the
first echo observation. Nevertheless, our analyses demonstrate the
wealth of information that can be extracted from observing dust scattered
x-ray echoes of bursts. Encouraged by these results, we provide guidelines
for optimal future observations of magnetar bursts, showing that improved
observations' timing, resolution and sensitivity would yield far better
constraints on both the distance to an SGR and the properties of the
intervening dust.

In order to evaluate the expected angular scattered width of the
ring we compare the effects of clouds width, exposure duration and
PSF. Substituting an SGR distance of 15$\,$kpc and $x=0.3$ (i.e.
main cloud at 4.5$\,$kpc) in Equation \ref{eq:geometry} we get a
spread of roughly 1$''$ for an assumed typical cloud width of
100$\,$pc, and a spread of $\Delta\theta=0.36''\left(t_{{\rm
start}}^{\frac{1}{2}}-t_{{\rm end}}^{\frac{1}{2}}\right)$ due to
exposure duration (where $t$ is the time passed in seconds since the
burst direct observation). This yields a width of $\sim$1$''$ for
the first XRT epoch, at a delay of 30 hours and duration of 40
minutes, due to the non-PSF effects. Thus the width is practically
set by the PSF of the telescope, with $\sim$20$''$ for ${\it
Swift}$/XRT vs. 1$''$ for $\emph{Chandra}$.

Effective constraining of $a_{{\rm max}}$ requires observations earlier
than 30 hours after the burst, and such a response time is suitable
for ${\it Swift}$/XRT. With a stronger scattered signal and a sharper
decay of the King component for earlier observations, resolving the
scattered signal from the source should be possible for a $\sim$1
hr observation taken 12 hours after the burst ($\theta_{{\rm obs}}=75''$).
Since observations should also cover the scattered flux power-law
regime, the burst fluence should be similar or larger than the 2006
August 6 burst.

\begin{figure}
\begin{centering}
\includegraphics[scale=0.45]{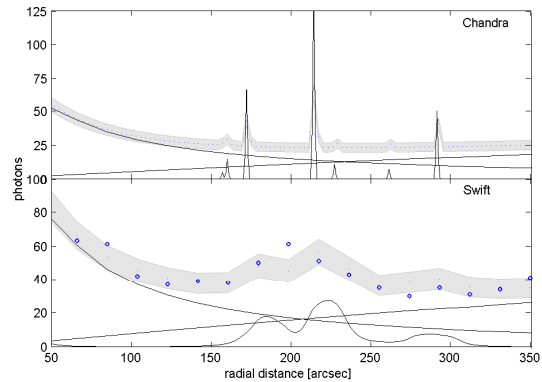}
\par
\end{centering}

\caption{\label{fig:chandra}A comparison of the best fit photon counts for
the second epoch 2.5\textminus{}4.5$\,$keV band with SGR at 16$\,$kpc,
${\it Swift}$ vs. $\emph{Chandra}$. Based on a $\emph{Chandra}$
4-fold signal and 3$''$ resolution. Contributions of clouds scattered
signal, background and fading halo are shown as continuous lines.
Poisson $1\sigma$ confidence limits for the best fit photon counts
are marked as gray filled area. ${\it Swift}$ actual measured photon
counts are marked in circles.}

\end{figure}

$\emph{Chandra}$ observation can be obtained with a response time
of a few days. The improved resolution and collection area of $\emph{Chandra}$
could significantly reduce errors and improve statistics. $\emph{Chandra}$
also enables a continuous observation due to its high orbit, thus
minimizing the spread due to exposure duration. A resolution of 2$''$\textminus{}3$''$
(accounting for PSF, exposure duration and cloud width) and a signal
roughly four times that of ${\it Swift}$ should improve the signal
to noise ratio, and enable resolving the location of individual clouds
along the line of sight. Each of the resolved clouds will provide
an independent constraint on the SGR distance. This will provide a
more accurate and, more importantly, a more robust distance estimate.
In Figure \ref{fig:chandra} we illustrate the advantages of using
$\emph{Chandra}$, comparing what $\emph{Chandra}$ would have seen
at the second XRT epoch to the actual ${\it Swift}$ observation.

The burst energy determines the proper time for the last effective
observation. A delay of 30 hours,
equivalent to $\theta=120''$, is located within the power-law regime
flux decay (Equation \ref{eq:approx1}), as seen in Figure \ref{fig:ML}.
Using our $q=-3.3$, combined with our measured scattered flux, we
get\begin{equation}
  F_{{\rm scat}}=
 F_0\,\frac{\Phi}{\Phi_{\rm \,August\,6,\,2006}}\left(\frac{t_{{\rm
delay}}}{30\,\mbox{hr}}\right)^{-1.85} \label{eq:lightCurve}
\end{equation}
where $F_0(2.5-4.5\,{\rm keV})=1.63 \times10^{-12}{\rm \,erg\,cm^{-2}\,s^{-1}}$ and $F_0(2.5-4.5\,{\rm keV})=4.45 \times10^{-12}{\rm \,erg\,cm^{-2}\,s^{-1}}$ are the fluxes from table \ref{tab:Best-fit-fluxes}. The equation is valid for $t_{{\rm delay}}\eqslantgtr30$ hours (power-law regime).
We calibrate the background photon count by the total background photon
count measured for ${\it Swift}$ (e.g. $0.14\,{\rm ph\,s^{-1}}$ for $2.5-4.5\,{\rm keV}$),
with the background per radial bin piling up as $\theta\propto\sqrt{t}$.
By requiring the bin which contains the largest fraction $f\le1$
of the scattered signal to be $N_{\sigma}$ times larger than the
background error we get a rough estimation of the maximum delay for
an effective observation:
\begin{equation}
\begin{array}{l}
  t= \\
t_0\left(\frac{\Phi}{\Phi_{\rm
\,August\,6,\,2006}}\frac{5}{N_{\sigma}}f\sqrt{\frac{10''}{\mbox{bin
width}}\frac{A_{{\rm eff}}}{A_{{\rm eff}}^{{\rm Swift}}}\frac{\Delta
t}{1\,\rm hr}}\right)^{0.48} \label{eq:lastObs}
\end{array}
\end{equation}

where $t_0(2.5-4.5\,{\rm keV})=135\,{\rm hr}$ and $t_0(4.5-8.5\,{\rm keV})=120\,{\rm hr}$, $A_{\rm eff}$ is the detector's effective area for the chosen energy band,
and $\Delta t$ is the observation's exposure time in hours.

For the ${\it Swift}$ case at $2.5-4.5\,{\rm keV}$ we get a $>5\sigma$ detection for the
first observation, at t=30$\,$hr, and a $\sim5\sigma$ detection
for the t=100$\,$hr second observation. A $\emph{Chandra}$ $A_{{\rm eff}}\approx4A_{{\rm eff}}^{{\rm Swift}}$,
$\Delta t=5\,$hr (for the latest observation), a bin width of 3$''$
and $f\approx1$ permits a $5\sigma$ detection as long as 370 hr
after a burst of the same fluence, thus probing dust grains down to
a size of $\sim0.022\,\mu\mbox{m}$ (vs. $\sim0.035\,\mu\mbox{m}$
in our case). Alternatively, a $\emph{Chandra}$ observation of $\Delta t=3\,$hr
after 100$\,$hr can yield a $5\sigma$ detection for a burst of a
fluence ten times weaker than that of 2006 August 6. Such SGR bursts
are more common, offering more opportunities for observations.

Equations \ref{eq:lightCurve} and \ref{eq:lastObs} quote the
fluence of the 2006 August 6 burst. The measured event fluence at the range 20\textminus{}200
keV is $\Phi=2.4\times10^{-4}\mbox{erg}\cdot\mbox{cm}^{-2}$ \citep{2006GCN..5426....1G}. We assume, at least for minor/intermediate bursts,
a similar spectrum across different energy outputs \citep[e.g.][]{2003AIPC..662...82O,2004ApJ...612..408F}. Hence, for events which fluence is given at 20\textminus{}200
keV, one should substitute the above cited fluence along with the new event fluence. For events where the band fluence (e.g. $2.5-4.5\,\mbox{keV}$) is more accessible, or events for which the similar spectrum assumption does not necessarily hold (e.g. giant flares), we supply an estimate of the 2006 August 6 burst fluence at the
$2.5-4.5\,\mbox{keV}$ and the $4.5-8.5\,\mbox{keV}$ energy bands.
\cite{2004ApJ...616.1148O}, \cite{2004ApJ...612..408F},
\cite{2007PASJ...59..653N} and \cite{2008ApJ...685.1114I} obtain
satisfying fits for the spectra of minor/intermediate SGR bursts
using models composed of two blackbody components, a harder
$kT\approx 9-10$ one and a softer $kT\approx 3-4$ one.
\cite{2008ApJ...685.1114I} compared three of the cited works and
found the bolometric luminosities of the two components to be
similar, with a possible saturation of the soft component above
$10^{41}\,{\rm erg\,s^{-1}}$. We adopt an equal luminosity for both
components. Calibrating all the cited models with the SGR fluence
per keV at $kT=20$ (derived from the total measured fluence and an
OTTB spectrum of $kT=20$ keV over the range of 20$-$200 keV), we
calculate the band fluence according to the blackbody temperatures
of each cited model, and estimate the fluence as the geometric mean
of the two extreme model results, obtaining
$\Phi_{2.5-4.5}=1.0^{+0.9}_{-0.4}\times 10^{-6}\,{\rm erg\,cm^{-2}}$
and $\Phi_{4.5-8.5}=8.3^{+5.0}_{-3.2}\times 10^{-6}\,{\rm
erg\,cm^{-2}}$. These values can be used for equations \ref{eq:lightCurve} and \ref{eq:lastObs} along with the respective band fluence of a new event, regardless of the event's spectrum.

A burst brighter than the one we analyzed should enable probing
smaller dust grains and even, if bright enough, allow the
determination of $a_{{\rm min}}$, the lower cutoff of the dust grain
size distribution. The launch of $\emph{NuSTAR}$ next year may
provide an opportunity to constrain $a_{{\rm min}}$ by observing
echoes from an intermediate burst, since for a given $a_{{\rm min}}$
the transition time from the power-law regime to the exponential
decay regime is proportional to $\theta^{2}$ and therefore to
$E^{-2}$ (Equation \ref{eq:typical angle}). $\emph{NuSTAR}$ has a
sensitivity range of $5-80\,{\rm keV}$ and a larger effective area
compared to $\emph{Chandra}$ (for $E<35\,{\rm keV}$ ). In addition,
the signal to noise at higher energies should improve due to the
spectrum of SGR bursts and to the decrease in the background level.
On the other hand, $\emph{NuSTAR's}$ relatively low resolution of
$43''$ will dilute the signal. Taking it all into account,
$\emph{NuSTAR}$ should be able to follow hard x-ray echoes from
intermediate bursts for a few days after the burst, thus probing
much smaller grains, and potentially identifying $a_{{\rm min}}$.

Generally, x-ray echoes observations taken during different bursts
of the same SGR can be combined to improve signal to noise ratio and
thus improve the constraints discussed in this work.

\section{Conclusions}

We used ${\it Swift}$/XRT observations following an intermediate
burst of SGR 1806\textminus{}20 to constrain the distance to the SGR
and to find the first x-ray echo constraints on molecular clouds'
dust properties. Based on the molecular clouds' properties along the
line of sight found by CE04 and the observed echoes of the burst we
constrain the distance with a lower limit of $9.4$$\,$kpc and an
upper limits of $18.6$$\,$kpc at a 90\% confidence level. This upper
limit can be considered the first direct upper limit set for the
distance to the SGR, as the upper limit set by
\citet{2005Natur.434.1112C} is questionable, and other distance
estimates are based on association rather than measured emission
from the source. Our distance constraints favor an energy output of
$\sim10^{46}\,$erg$\,$s$^{-1}$ for the 2004 giant flare of SGR
1806\textminus{}20, leaving Galactic vs. extragalactic giant bursts
rates possible tension and the lack of post-burst emission features
changes in SGR 1806\textminus{}20 as open issues.

We introduce the use of observations of dust x-ray echoes for
probing the dust properties of molecular clouds. Fitting the
spectral and temporal signal evolution using a dust scattering model
with an assumed power-law dust grain size distribution we find a
power-law index of $q=-3.3_{-0.7}^{+0.6}$ ($1\sigma$) and that the
dust maximal grain size $a_{{\rm max}}>0.1\,\mu\mbox{m}$
($2\sigma$). These results are of special interest since the dust
along the line of sight to SGR 1806\textminus{}20 resides in
molecular clouds, which dust properties have been poorly explored.
The constraints we got, imply that the dust grain size distribution
in molecular clouds may be similar to the one found in diffused ISM
(e.g. \citealt{1977ApJ...217..425M}).

The wealth of data obtained using ${\it Swift}$/XRT encourage us to
suggest the use of future x-ray echoes observations. For an
intermediate SGR 1806\textminus{}20 burst, a ${\it Swift}$
observation starting half a day after the burst and a
$\emph{Chandra}$ observation a few days after the burst should yield
superior dust grain size distribution characterization and SGR
distance constraints, respectively. A $\emph{Chandra}$ observation
should enable, in addition, a quality mapping of clouds' locations
along the line of sight. $\emph{NuSTAR}$, planned to be launched
next year, could potentially probe $a_{{\rm min}}$, the minimal dust
grain size.

The authors would like to thank Bruce Draine, Eli Dwek, Derek Fox
and Amiel Sternberg for helpful discussions. Special thanks to
Chryssa Kouveliotou for helpful comments and careful reading of the
manuscript. This work was partially supported by the Israel Science
Foundation (grant No. 174/08) and by an IRG grant. EOO is supported
by an Einstein fellowship and NASA grants.

%\bibliographystyle{mn2e}
%\bibliography{echoes,spec04}

\end{document}